\documentclass[aps,rmp,nofootinbib]{revtex4}
\usepackage{epsf}
\begin{document}
\title{To see a world in a grain of sand}
\author{Shou-Cheng Zhang}
\affiliation{Department of Physics, Stanford University, Stanford, CA 94305}

\begin{abstract}
Throughout John Wheeler's career, he wrestled with big issues like
the fundamental length, the black hole and the unification of quantum
mechanics and relativity. In this essay, I argue that solid
state physics -- historically the study of silicon,
semiconductors and sand grains -- can give
surprisingly deep insights into the big questions of the world.
\end{abstract}
\maketitle

\tableofcontents

\section{INTRODUCTION}

\label{sec:introduction}
Modern physics is built upon three principal
pillars, quantum mechanics, special and general relativity.
Historically, these principles were developed as logically
independent extensions of classical Newtonian mechanics.
While each theory constitutes a logically
self-consistent framework, unification of these fundamental
principles encountered unprecedented difficulties. Quantum mechanics and special
relativity were unified in the middle of the last century,
giving birth to relativistic quantum field theory. While
tremendously successful in explaining experimental
data, ultraviolet infinities in the calculations hint that the
theory can not be in its final form. Unification of quantum mechanics with
general relativity proves to be a much more difficult task and
is still the greatest unsolved problem in theoretical physics.

In view of the difficulties involved with unifying these principles,
we can ask a simple but rather bold question:
Is it possible that the three principles are not
logically independent, but rather there is a hierarchical order in their logical
dependence? In particular, we notice that both relativity principles can
be formulated as statements of symmetry. When applying non-relativistic
quantum mechanics to systems with a large number of degrees of freedom,
we sometimes find that symmetries can emerge in the low energy sector, which
are not present in the starting Hamiltonian. Therefore, there is a logical
possibility that one could start from a single non-relativistic Schr\"odinger
equation for a quantum many-body problem, and discover relativity principles
emerging in the low energy sector. If this program can indeed be realized,
a grand synthesis of fundamental physics can be achieved. Since non-relativistic
quantum mechanics is a finite and logically self-consistent framework,
everything derived from it should be finite and logically consistent as well.

The Standard Model in particle physics is described by a relativistic quantum
field theory and is experimentally verified below the energy scale of $10^{3} GeV$.
On the other hand, the Planck energy scale, where quantum gravitational
force becomes important, is at $10^{19} GeV$. Therefore, we need to extrapolate
$16$ orders of magnitude to guess the new physics beyond the standard framework
of relativistic quantum field theory. It is quite conceivable that Einstein's
principle of relativity is not valid at Planck's energy scale, it could emerge
at energies much lower compared to the Planck's energy scale through the magic
of renormalization group flow. This situation is analogous to one in condensed matter
physics, which deals with phenomena at much lower absolute energy scales. The
``basic" laws of condensed matter physics are well-known at the Coulomb
energy scale of $1\sim 10 eV$; almost all condensed matter
systems can be well described by a non-relativistic Hamiltonian of the
electrons and the nuclei\cite{laughlin2000A}.
However, this model Hamiltonian is rather inadequate
to describe the various emergent phenomena, like superconductivity, superfluidity,
the quantum Hall effect (QHE) and magnetism, which all occur at much lower energy scales,
typically of the order of $1 meV$. These systems are best described by
``effective quantum field theories", not of the original electrons, but
of the quasi-particles and collective excitations. In this lecture, I shall
give many examples where these ``effective quantum field theories" are
relativistic quantum field theories or topological quantum field theories,
bearing great resemblance to the Standard Model of
elementary particles. The collective behavior of many strongly interacting
degrees of freedom is responsible for these striking emergent phenomena.
The laws governing the quasi-particles and the collective excitations are
very different from the laws governing the original electrons and
nuclei\cite{anderson1972A}.
This observation inspires us to construct models of elementary particles by
conceptually visualizing them as quasi-particles or collective excitations
of a quantum many-body system, whose basic constituents are governed by a
simple non-relativistic Hamiltonian. This point of view is best summarized
by the following diagram:

\begin{center}
\begin{tabular}{ccc}
Planck energy at $10^{19} GeV$ & $\Leftrightarrow$ &   Coulomb energy at $10 eV$  \\
$\uparrow$ ?     & &   $\downarrow$ \\
Standard Model at $10^{3} GeV$& $\Leftrightarrow$ &   Superconductivity, QHE, Magnetism etc at 1 meV \\
Relativistic quantum field theory of elementary particles & &
Effective quantum field theory of quasi-particles
\end{tabular}
\end{center}

The conceptual similarity between particle physics and
condensed matter physics has played a very important role in the history
of physics. A crucial ingredient of the Standard Model, the idea
of spontaneously broken symmetry and the Higgs mechanism, first
originated from the BCS theory of superconductivity. This example
vividly shows that the physical vacuum is not empty, but a condensed
state of many interacting degrees of freedom. Another fundamental
concept is the idea of renormalization group transformation, which
was simultaneously developed in the context of particle physics and
in the study of critical phenomena. From the theory of renormalization
group, we learned that symmetries can emerge at the low energy sector,
without being postulated at the microscopic level. Today, as physicists
face unprecedented challenges of unifying quantum mechanics with relativity,
and tackling the problem of quantum gravity, it is useful to
look at these historic successes for inspiration. A new
era of close interaction between condensed matter physics and
particle physics could shed light on these grand challenges of
theoretical physics.

\section{EXAMPLES OF EMERGENCE IN CONDENSED MATTER SYSTEMS}
\label{emergence}

In this section, we review some well-known examples in condensed
matter physics, where one starts from a quantum many-body system
at high energies and arrives at a relativistic or topological field
theory of the low energy quasi-particles and elementary
excitations. The high energy models often look simple and
innocuous, yet the emergent low energy phenomena and their
effective field theory description are profound and beautiful.

\subsection{2+1 dimensional QED from superfluid helium films}
\label{helium}

\begin{figure}[h]
\centerline{\epsfysize=4cm \epsfbox{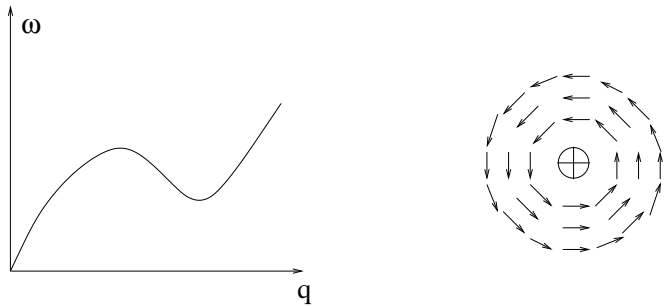} }
\caption{Collective excitations of a neutral 2D superfluid film are the
sound waves and the vortices. In the long wave length limit, the sound
wave maps onto the Maxwell fields, while vortices map onto electric
charges.
}
\label{vortex}
\end{figure}

Let us first start from the physics of a superfluid film. The mean
velocity of the helium atoms are
significantly lower compared to the speed of light, therefore,
relativistic effects of the atoms can be completely neglected. The
basic non-relativistic Hamiltonian for this system of identical bosons
can be expressed in the following close form:
\begin{eqnarray}
H = \frac{1}{2m} \sum_n \vec{p}_n^2 + \sum_{n < n'} V(x_n - x_{n'})
\label{helium_H}
\end{eqnarray}
where $V$ is the inter-atomic potential, whose form depends on the
details of the system. However, for a large class of generic
interaction potentials, the system flows towards a universal low
energy attractive fix point, namely the superfluid ground state.
At typical inter-atomic energy scales of a few $eV$'s,
helium atoms are the correct dynamic variables, and the
Hamiltonian (\ref{helium_H}) is the correct model Hamiltonian.
However, at the energy scale characteristic of the
superfluid transition, which is of the order of $1K\sim 10^{-4}
eV$, the correct dynamical variables are sound wave modes
and the vortices of the superfluid film. (See fig. (\ref{vortex}). for an
illustration).

The remarkable thing is that the effective field theory model for
these low energy degrees of freedom is exactly the relativistic
quantum electrodynamics (QED) in $2+1$ dimensions! This connection
was established by the work of Ambegaokar, Halperin, Nelson and
Siggia\cite{ambegaokar1980A} and derived from the point of
view of vortex duality\cite{fisher1989A}. To see how
this works, let us recall that the basic hydrodynamical variables
of the superfluid film are the density $\rho(x)$ and the velocity
$v_i(x)$ fields, ($i=1,2$), satisfying the equation of continuity
\begin{eqnarray}
\partial_t \rho + \partial_i (\bar\rho v_i) =0
\end{eqnarray}
where $\bar\rho$ is the average density of the fluid.
Now let us recall that in $2+1$ dimensions, the electric field
$E_i$ has two components while the magnetic field $B$ has only one
component, which can therefore be identified as a scalar.
Faraday's law of induction is given by the Maxwell
equation:
\begin{eqnarray}
\frac{1}{c}\partial_t B + \epsilon_{ij} \partial_i E_j =0
\label{faraday}
\end{eqnarray}
where $\epsilon_{ij}$ is the antisymmetric tensor in two dimensions.
Therefore, if we make the following identification,
\begin{eqnarray}
B \ \ \Leftrightarrow \ \ -c\frac{\rho}{\bar\rho} \ \ \ \ \ \ \ \ \ E_i \ \
\Leftrightarrow \ \ \epsilon_{ij} v_j
\end{eqnarray}
we see that the equation of continuity of the superfluid film
agrees exactly with Faraday's law as expressed in the Maxwell's
equation (\ref{faraday}).
Next we examine the fluid velocity in the presence of a
vortex, with unit vorticity, located at the position $x_n$.
The superfluid state has a well defined $U(1)$ order parameter,
and the velocity field can be expressed in terms of the phase,
$\phi$, of the $U(1)$ order parameter:
\begin{eqnarray}
v_i = \frac{\hbar}{m}\partial_i \phi
\label{1stJ}
\end{eqnarray}
Because of the single valuedness of the quantum mechanical wave
function, $e^{i\phi}$ must be single valued. Therefore,
the superflow around a vortex is {\it quantized}:
\begin{eqnarray}
\int \vec v \cdot d\vec l = 2\pi \frac{\hbar}{m} q
\end{eqnarray}
where $q$ is an integer. For elementary vortices, $q=\pm 1$.
The differential form of this integral equation is
\begin{eqnarray}
\epsilon_{ij} \partial_i v_j = 2 \pi \rho_v
(x)
\label{flow}
\end{eqnarray}
where $\rho_v(x)=\sum_n q_n \delta(x-x_n)$ is the density of the
vortices and $q_n=\pm 1$ is the vorticity. If we identify the vortex
density with the electric charge density in Maxwell's equations, we
see that equation (\ref{flow}) is nothing but Gauss's law in
$2+1$ dimensions:
\begin{eqnarray}
\partial_i E_i = 2\pi \rho_v (x)
\label{Gauss}
\end{eqnarray}
Next let us investigate the dynamics of the superfluid velocity
$v_i$, through the Josephson equations of superfluidity. The
first Josephson equation relates the superfluid velocity to the
gradient of the superfluid phase, $\phi$, as expressed in
(\ref{1stJ}). The second Josephson equation relates the
time derivative of the phase to the chemical potential $\hbar\partial_t \phi =-\mu$.
Combining the two Josephson equations, we obtain,
\begin{eqnarray}
\partial_t v_i = \frac{\hbar}{m} \partial_t \partial_i \phi = -\frac{1}{m} \partial_i \mu =
- \frac{\kappa}{m\bar\rho}  \partial_i \rho
\end{eqnarray}
where we use the compressibility $\kappa=\bar\rho \frac{\partial\mu}{\partial\rho}$
to express the chemical
potential $\mu$ in terms of the density $\rho$. This equation agrees
exactly with the source-free Maxwell equation
\begin{eqnarray}
c \epsilon_{ij} \partial_j B = \partial_t E_i
\label{source_free}
\end{eqnarray}
provided one identifies the speed of light as $c^2=\kappa/m$.
This equation
needs to be modified in the presence of the vortex flow $J_i^v$,
which unwinds the $U(1)$ phase by $2\pi$ each time a vortex passes.
The vortex current satisfies the equation of continuity
\begin{eqnarray}
\partial_t \rho_v + \partial_i J^v_i =0
\label{v_continuity}
\end{eqnarray}
Therefore, the source free Maxwell equation (\ref{source_free})
acquires a additional term, in order to be compatible with
both (\ref{v_continuity}) and (\ref{Gauss}):
\begin{eqnarray}
c \epsilon_{ij} \partial_j B = \partial_t E_i +2\pi J^v_i
\label{with_source}
\end{eqnarray}
This is nothing but Ampere's law, supplemented by Maxwell's
displacement current.

This proves the complete equivalence between the superfluid
equations and Maxwell's equations in $2+1$ dimensions.
Interestingly enough, we seem to have completed a rather curious loop.
Starting from the relativistic Standard Model of the quarks and
leptons, one arrives at an effective non-relativistic model of
the helium atoms (\ref{helium_H}). However, as one reduces the
energy scale further, the effective low energy degrees of freedom
become the sound modes and the vortices, which are described by the
field theory of $2+1$ dimensional quantum electrodynamics, very similar to
the model we started from in the first place! A ``civilization"
living inside the helium film would first discover the
Maxwell's equations, and then, after much harder work, they would
establish equation (\ref{helium_H}) as their ``theory of everything".

Superfluid $^4He$ films are relatively simple because the $^4He$
atom is a boson. The superfluidity of $^3He$ is much
more complex, with many competing superfluid phases. In fact,
Volovik\cite{volovik2001A} has pointed out that many phenomena
of the superfluid phase of $^3He$ share striking similarities with the
Standard Model of elementary particles. These similarities
inspired him to pioneer a program to address cosmological
questions by condensed matter analogs.

\subsection{Dirac fermions of $d$ wave superconductors}
\label{Dirac}
Having considered the low energy properties of a superfluid,
let us now consider the low energy excitations of a superconductor,
with $d$ wave pairing symmetry. In this case, there are low
energy fermionic excitations besides the bosonic excitations.
This system is realized in the high $T_c$ superconductors.
The microscopic Hamiltonian is the two dimensional (2D) Hubbard model, or the $t-J$ model,
expressible as
\begin{eqnarray}
H = -t \sum_{\langle ij\rangle,\sigma} c^\dagger_{i\sigma} c_{j\sigma}
+ J \sum_{\langle ij\rangle} \vec S_i \cdot \vec S_j
\label{t-J}
\end{eqnarray}
where $c^\dagger_{i\sigma}$ is the electron creation operator
on site $i$ with spin $\sigma$, $\vec S_i$ is the electron
spin operator and $\langle ij\rangle$ denotes the nearest
neighbor bond on a square lattice. Double occupancy of a
single lattice site is forbidden.

This model is valid at the energy scale of $150 meV$, which is the typical
energy scale of the antiferromagnetic exchange $J$. When the filling factor
$x$ lies between $10\%$ and $20\%$, the ground state of this model is
believed to be a $d$ wave superconductor. There is indeed overwhelming
experimental evidence that the pairing symmetry of the high $T_c$ superconductor
is $d$ wave like. Remarkably, the elementary excitations in this case
can be described by the $2+1$ dimensional Dirac Hamiltonian.
In contrast to the $t-J$ model, which is valid at the energy scale of
$100meV$, the effective Dirac Hamiltonian for the $d$ wave quasi-particles is
valid at much lower energy, typically of the order of $30meV$, which is
the maximal gap. While the connection between the $t-J$ model and
$d$ wave superconductivity still needs to be firmed established, the
connection between the $d$ wave BCS quasi-particle Hamiltonian and the
Dirac equation is well-known in the condensed
matter community\cite{volovik1993A,simon1997A,balents1998A,franz2002A}.
Here we follow a pedagogical presentation by Balents,
Fisher and Nayak\cite{balents1998A}.

The BCS mean field Hamiltonian for a $d$ wave superconductor is given
by
\begin{equation}
H = \sum_{k \alpha} \epsilon_k c_{k\alpha}^\dagger c_{k \alpha} +
\sum_k [ \Delta_k c_{k \uparrow}^\dagger c_{-k\downarrow}^\dagger
+ \Delta_k^* c_{-k\downarrow} c_{k\uparrow} ]    .
\end{equation}
where $\epsilon_k$ is the quasi-particle dispersion relation, and
$\Delta_k$ is the $d$ wave pairing gap, given by
\begin{equation}
\epsilon_k = -2t (cos k_x + cos k_y) \ \ , \ \Delta_k = \Delta_0 (cos k_x - cos k_y)
\end{equation}
One can introduce a four component spinor
\begin{equation}
\Upsilon_{a\alpha}(\vec{k}) = \left[ \begin{array}{c}
\Upsilon_{11} \\ \Upsilon_{21} \\ \Upsilon_{12} \\ \Upsilon_{22}
\end{array} \right] = \left[ \begin{array}{c}
c_{k\uparrow}^{\vphantom\dagger} \\ c_{-k\downarrow}^{\dagger} \\
c_{k\downarrow}^{\vphantom\dagger} \\ -c_{-k\uparrow}^{\dagger}
\end{array}\right].
\end{equation}
which doubles the number of degrees of freedom. This can be compensated
by summing over only half of the Brillouin zone, say $k_y>0$.
In terms of these variables, the BCS Hamiltonian becomes
\begin{equation}
H = \left. \sum_{k,k_y>0} \right. \Upsilon^\dagger_{a\alpha}(\vec{k})
[ \tau^z \epsilon_k + \tau^+ \Delta_k
+ \tau^- \Delta^*_k ]_{ab} \Upsilon_{b\alpha}(\vec{k})  ,
\end{equation}
where $\vec{\tau}_{ab}$ are the standard Pauli matrices
acting in the particle/hole subspace.

The d-wave nodes are approximately located
near the special wave vectors
$\vec{K}_{1} = (\pi/2,\pi/2)$,
$\vec{K}_{2} = (-\pi/2,\pi/2)$, $\vec{K}_{3} = -\vec{K}_{1}$ and
$\vec{K}_{4} = -\vec{K}_{2}$. In order to obtain a long wave length
and low energy description, we can expand around the nodal points
$\vec{K}_{1}$ and $\vec{K}_{2}$, which satisfy the $k_y>0$ constraint.
The nodal points $\vec{K}_{3}$ and $\vec{K}_{4}$ are automatically
taken into account in the $\Upsilon$ spinor.

\begin{figure}[h]
\centerline{\epsfysize=5cm \epsfbox{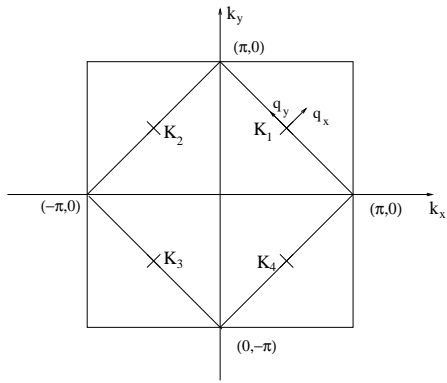} }
\caption{A 2D $d$ wave superconductor has four nodes, indicated by
$K_1, K_2, K_3$ and $K_4$. Around these nodal points, BCS quasi-particles
obey the massless Dirac equation.
}
\label{nodes}
\end{figure}

Introducing the
rotated coordinates $q_x$ and $q_y$, as indicated in fig. (\ref{nodes}),
and the effective spinors
\begin{equation}
\Psi_{1a\alpha}(\vec{q}) = \Upsilon_{a\alpha}(\vec{K}_{1}+\vec{q}) \ \ , \ \
\Psi_{2a\alpha}(\vec{q}) = \Upsilon_{a\alpha}(\vec{K}_{2}+\vec{q})
\end{equation}
we obtain
\begin{equation}
H = \sum_{q\in K_1} \Psi^\dagger_{1a\alpha}(\vec{q})
[ \tau^z \epsilon_{K_1+q} + \tau^+ \Delta_{K_1+q}
+ \tau^- \Delta^*_{K_1+q} ]_{ab} \Psi_{1b\alpha}(\vec{q}) + (1\leftrightarrow 2)
\end{equation}
Here ${q\in K_1}$ denotes a momentum sum near the vector $K_1$.
Expansion near $K_1$ gives
\begin{equation}
\epsilon_{K_1+q} \approx v_F q_x \ \ , \ \ \Delta_{K_1+q} \approx \Delta q_y
\end{equation}
A similar expansion applies for $K_2$. Going to the continuum limit, we obtain
the Hamiltonian density
\begin{eqnarray}
  {\cal H} & = & \Psi_{1a\alpha}^\dagger [v_F\tau^z i \partial_x +
  (\tilde{\Delta} \tau^+ + \tilde{\Delta}^* \tau^-) i \partial_y ]_{ab} \Psi_{1b\alpha}
  \nonumber \\
  & & + (1 \leftrightarrow 2; x \leftrightarrow y)  ,
\end{eqnarray}
which is exactly the Dirac Hamiltonian density in $2+1$ dimensions.
Once again, we see the emergent
relativistic behavior of a quantum many-body system. We start
from a non-relativistic interacting fermion problem at higher energies, but
recover a relativistic Dirac equation at low energies.

\subsection{Emergence of a topological quantum field theory}
\label{topology}
When Einstein first wrote down his field equation of general relativity,
he said that the left-hand side of the equation that had to do specifically
with geometry and gravity was beautiful - it was as if made of marble.
But the right-hand side of the equation that had to do with matter and how
matter produces gravity was ugly - it was as if made of wood.
Taking Einstein's aesthetic point of view one step further, one is tempted to
construct a fundamental theory by starting with the description of
the topology, or a topological field theory without matter and without
even geometry from the start. Having demonstrated that the relativistic
Maxwell equation and Dirac equation can emerge in the low energy sector
of a quantum many-body problem, I now give an example demonstrating how
a topological quantum field theory, namely the Chern-Simons (CS) theory,
can emerge from the matter degrees of freedom in the low energy sector
of the QHE. The CS topological
quantum field theory was derived microscopically by Zhang, Hansson and
Kivelson\cite{zhang1989A}, and reviewed extensively in ref. \cite{zhang1992A}.

The basic Hamiltonian of QHE is simply that of a two-dimensional
electron gas in a perpendicular magnetic field.
\begin{eqnarray}
H =
 \frac{1}{2m} \sum_n \left[\vec{p}_n - \frac{e}{c}
  \vec{A}(x_n)\right]^2
  + \sum_n e A_0 (x_n) + \sum_{n < n'} V(x_n - x_{n'})
\label{qhe}
\end{eqnarray}
where $\vec{A}$ is the vector potential of the external magnetic field,
which in the symmetric gauge can be expressed as
\begin{equation}
A_i = \frac{1}{2} B \epsilon_{ij} x_j
\end{equation}
$A_0$ is the scalar potential of the external electric field,
$E_i = - \partial_i A_0$,
and $V(x)$ is the interaction between the electrons. For high magnetic
fields, the electron spins are polarized along the same direction.
Since the spin wave function is totally symmetric, the Hamiltonian
(\ref{qhe}) operates on orbital wave functions which are totally
antisymmetric.
This model is valid at the Coulombic energy scale of several $eV$'s and has
no particular symmetry or topological properties. Since the external magnetic
field breaks time reversal symmetry, an invariant tensor $\epsilon_{ij}$ can
be introduced into the response function, and in particular, one can have
a current $J_i$, which flows transverse to the applied electric field $E_j$,
given by
\begin{eqnarray}
J_i = \rho_H^{-1} \epsilon_{ij} E_j
\label{Hall}
\end{eqnarray}
where $\rho_H$ is defined as the Hall resistance.
Since the electric field is perpendicular to the induced current, it does no
work on the electrons, and the current flow is dissipationless. The 2D
electron density $n$ in a magnetic field $B$ is best measured in terms of a
dimensionless quantity called the filling factor $\nu=n/n_B$, where
$n_B=B/\phi_0=eB/hc$ is the magnetic flux density. QHE is the
remarkable fact that the coefficient of the Hall response is quantized, given by
\begin{eqnarray}
\rho_H = \nu^{-1} \frac{h}{e^2}
\label{quantized}
\end{eqnarray}
when the filling fraction is near a rational number $\nu=p/q$ with odd denominator
$q$. QHE at fractional values of $\nu$ is referred to as the fractional QHE (FQHE).

FQHE is described by Laughlin's celebrated wave function.
There is also an alternative way to understand this profound effect by the
Chern-Simons-Landau-Ginzburg (CSLG) effective field theory\cite{zhang1992A}.
The idea is to perform a singular gauge transformation on (\ref{qhe}),
and turn electrons into bosons. This is only possible in $2+1$ dimensions.
Consider another Hamiltonian
\begin{equation}
H' =
 \frac{1}{2m} \sum_n \left[\vec{p}_n - \frac{e}{c}
  \vec{A}(x_n) - \frac{e}{c} \; \vec{a}(x_n) \right]^2
  + \sum_n e A_0 (x_n) + \sum_{n < n'} V(x_n - x_{n'})
\label{qhe'}
\end{equation}
Every symbol in $H'$ has the same meaning as in $H$, except the new
vector potential $\vec{a}$, which describes a {\em gauge interaction}
among the particles and is given by
\begin{equation}
\vec{a}(x_n) = \frac{\phi_0}{2\pi} \frac{\theta}{\pi} \sum_{{n'}\neq n}
  \vec{\nabla} \; \alpha_{n{n'}}
\label{statistical}
\end{equation}
where $\phi_0 = hc/e$ is the unit of flux quantum
and $\alpha_{n{n'}}$ is the angle
sustained by the vector connecting particles $n$ and $n'$ with an
arbitrary vector specifying a reference direction, say the $\hat{x}$
axis. The crucial difference here is while $H$ operates on a fully
antisymmetric fermionic wave function, $H'$ operates on a fully
symmetric bosonic wave function. One can prove an exact theorem
which states that these two quantum eigenvalue problems are
equivalent to each other when $\theta/\pi=(2k+1)$ is an odd integer.
In this case, each electron is attached to an odd number of
fictitious quanta of gauge flux (cause by $a$), so that their exchange
statistics in $2+1$ dimensions becomes bosonic.
These bosons, called composite bosons\cite{girvin1987A,zhang1989A,read1989A},
see two different types of gauge fields: the external magnetic
field $A$, and an internal statistical gauge field $a$. The average of
the internal statistical gauge field depends on the density of the
electrons. When the external magnetic field is such that the filling
fraction $\nu=n_B/n=1/2k+1$ is the inverse of an odd integer, we can always
choose $\theta=(2k+1)\pi$ so that the net field seen by the composite
bosons cancels each other on the average.

\begin{figure}
\centerline{\epsfysize=6cm \epsfbox{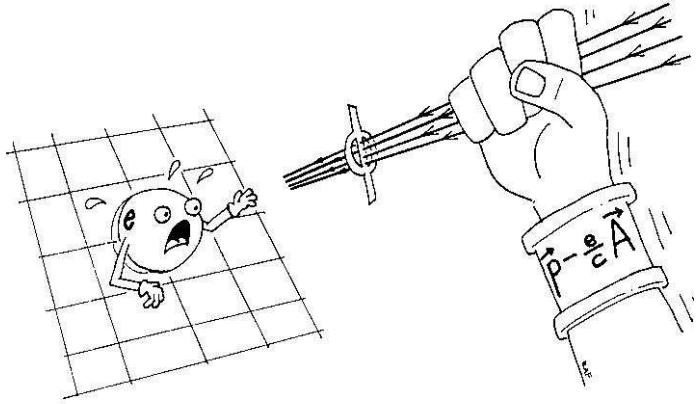} }
\caption{An electron just before the flux transmutation operation.
(taken from the PhD thesis of D. Arovas, illustrated by Dr. Roger Freedman).
}
\label{flux}
\end{figure}

The statistical transmutation from electrons to composite bosons can be
naturally implemented in quantum field theory through the Cherm-Simons term.
The Chern-Simons Lagrangian is given by
\begin{equation}
{\cal L} = \frac{1}{2} \left(\frac{\pi}{\theta} \right)
  \frac{1}{\phi_0} \; \varepsilon^{\mu v\rho} \;
  a_\mu \; \partial_v \; a_\rho - a_\mu \; j^\mu
\label{cslg}
\end{equation}
here $j^\mu$ is the current of the composite boson, and $\mu=0,1,2$ is the
space-time index in $2+1$ dimensions. The equation of motion for the
$a_0$ field is
\begin{equation}
\epsilon^{ij} \; \partial_i \;
  a_j (x) = \phi_0 \; \frac{\theta}{\pi} \; \rho(x)
\end{equation}
whose solution for $\rho(x)=\sum_n \delta(x-x_n)$ exactly gives the
statistical gauge field in (\ref{statistical}).

Now we can present the key argument of the CSLG theory\cite{zhang1992A}
of QHE. Even though
course the statistical transformation can be performed in any
$2+1$ dimensional systems, this does not mean that the low energy
limit of any $2+1$ dimensional system is given by a CS theory, since
the partition function also involves the integration over the matter
fields $j^\mu$  in the second term of (\ref{cslg}).  The
key observation is that at the special filling factors of $\nu=1/2k+1$,
the combined external and statistical magnetic field seen by the composite
boson vanishes, therefore, composite bosons naturally condense into a
superfluid state. This is the ``magic" of the magic filling factors
$\nu=1/2k+1$. We already showed in section (\ref{helium}) that the effective
field theory of a $2+1$ dimensional bosonic condensate is the $2+1$ dimensional
Maxwell theory. Therefore, the integration over the matter fields in
(\ref{cslg}) gives the Maxwell Lagrangian, $f^2_{\mu\nu}$. In $2+1$ dimensions,
the CS term contains one fewer derivative compared with the Maxwell
term, it therefore dominates in the long-wave length and low-energy limit.
Therefore, the effective Hamiltonian of FQHE is just the topological
CS theory, without the matter current term in (\ref{cslg}).

Matter degrees of freedom in the starting Hamiltonian (\ref{qhe})
are magically turned into topological degrees of freedom of the CS
field theory. Alchemy works! Wood is turned into marble!
Many people argued that a quantum
theory of gravity should be formulated independent of the
background metric. The emergent CSLG theory starts from matter
degrees of freedom in a background setting, but the resulting
effective field theory is independent of the background metric.
This demonstrates that in principle, background independent
theory can indeed be constructed from non-relativistic quantum
many-body systems. In fact, the CSLG theory also leads to
a beautiful duality symmetry based on the discrete $SL(2,Z)$ group,
very similar to the duality symmetry in the 4D Seiberg-Witten
theory.
This duality symmetry is again emergent, and it predicts the global
phase diagram of the QH Hall system. The phase diagram has a
beautiful fractal structure, with one phase nested inside each
other, iterated {\it ad infinitum}\cite{kivelson1992A}.

\section{THE FOUR DIMENSIONAL QUANTUM HALL EFFECT}
\label{4dqhe}

In the previous sections we saw that the collective behavior of
quantum many-body systems often gives rise to novel emergent
phenomena in the low energy sector, which are described in
terms of relativistic or topological quantum field theories.
Therefore, one can't help but wonder if the Standard Model could
also work this way. The problem is that the well-understood
examples of emergent relativistic behaviors in quantum many-body
systems work only for lower dimensions, and these models do
not have sufficient richness yet. In order for the Stanford Model
to appear as emergent behavior, we are led to study higher-dimensional
quantum many-body systems, specially the higher-dimensional
generalizations of QHE.

\subsection{The model}

Of all the novel quantum many-body systems, QHE
plays a very special role: it is the only well understood condensed
matter system whose low energy limit is a topological quantum field
theory. Unlike most other emergent phenomena, like superconductivity and
magnetism, QHE works only in two spatial dimensions. There are
various ways to see this. First of all, the Hall current is non-dissipative.
For the electric field to do no work on the current, the
current must flow in a direction perpendicular to the direction
of the electric field. In two spatial dimensions, given the direction
of the electric field, there is an unique transverse direction for the
Hall current, given by equation (\ref{Hall}).
Since the current and the electric field both
carry spatial vector indices, the response must therefore be a rank-two
tensor. But there are no natural rank-two antisymmetric tensors in
higher dimensions! Secondly,
both the single-particle wave function and Laughlin's many-body wave function
make extensive use of complex coordinates of particles,
which can only be done in two spatial dimensions.
This suggests that the higher-dimensional generalization of QHE would
necessarily involve a higher-dimensional generalization of complex
numbers and analytic functions.
In fact, both of these considerations lead to the same higher-dimensional
structure, as we shall explain below.

In higher dimensions, given a direction of the electric field, there is
no unique transverse direction for the Hall current to flow. However,
this statement holds only if we consider the $U(1)$ charge current. If the
underlying particles -- and the associated currents -- carry a non-abelian,
{\it e.g.} $SU(2)$ quantum number, an unique prescription for the current
can be given in {\it four} dimensions. In four dimensions, given a fixed
direction of the electric field, say along the $x_4$ direction, there are
three transverse directions. If the current carries a $SU(2)$ isospin label,
it also has three internal isospin directions. In this case, the current
can flow exactly along the direction in which the isospin is pointing. In this
prescription, no preferential direction in space or isospin is picked. The
system is invariant under a {\it combined} rotation of space and isospin.
To be more precise, the mathematical generalization of (\ref{Hall})
in four dimensions is
\begin{equation}
J_\mu^i = \sigma \eta^i_{\mu\nu} E_\nu
\label{4dHall}
\end{equation}
Here $\sigma$ is the generalized Hall conductivity,
$\eta^i_{\mu\nu}$ is the t' Hooft tensor, explicitly given by
$\eta_{\mu\nu}^i=\epsilon_{i\mu\nu4}+\delta_{i\mu}\delta_{4\nu}-
\delta_{i\nu}\delta_{4\mu}$ and $J_\mu^i$ is the
isospin current and $E_\nu$ is the electric field.
Here $\mu,\nu=1,2,3,4$ label the spatial
directions and $i=1,2,3$ label the isospin directions.
From (\ref{4dHall}),
we see easily that if $E_\nu$ points along the $x_4$ direction, the current
flows along the $x_{1,2,3}$ directions, explicitly determined by the
direction of the isospin. Therefore, the t' Hooft tensor is exactly the
rank-two antisymmetric tensor we were looking for!
The occurrence of the t' Hooft tensor suggests that
this problem must have something to do with the $SU(2)$
instanton\cite{belavin1975A}, where the
t' Hooft tensor was first introduced. It is not only an {\it invariant}
tensor under combined spatial and isospin rotations, it also satisfies
a self-duality condition:
\begin{equation}
\eta^i_{\mu\nu} = \epsilon_{\mu\nu\rho\lambda} \eta^i_{\rho\lambda}
\label{self-dual}
\end{equation}
Self-duality and anti-self-duality are the hallmarks of the $SU(2)$
Yang-Mills instanton.

Now let us motivate the problem from the point of view of generalizing
complex numbers. The natural generalizations of complex numbers are
quaternionic numbers, first discovered by Hamilton. A quaternionic
number is expressed as $q=q_0+q_1 i + q_2 j + q_3 k$, where $i,j,k$ are
the three imaginary units. This again suggests that the most natural
generalization of QHE is from 2D to 4D, where quaternionic numbers can be interpreted
as the coordinates of particles in four dimensions. Unlike complex numbers,
quaternionic numbers do not commute with each other. In fact, the three
imaginary units of quaternionic numbers can be identified with the three
generators of the $SU(2)$ group. This suggests that the underlying
quantum mechanics problem should involve a non-abelian $SU(2)$ gauge field.

\begin{figure}[h]
\centerline{\epsfysize=4cm \epsfbox{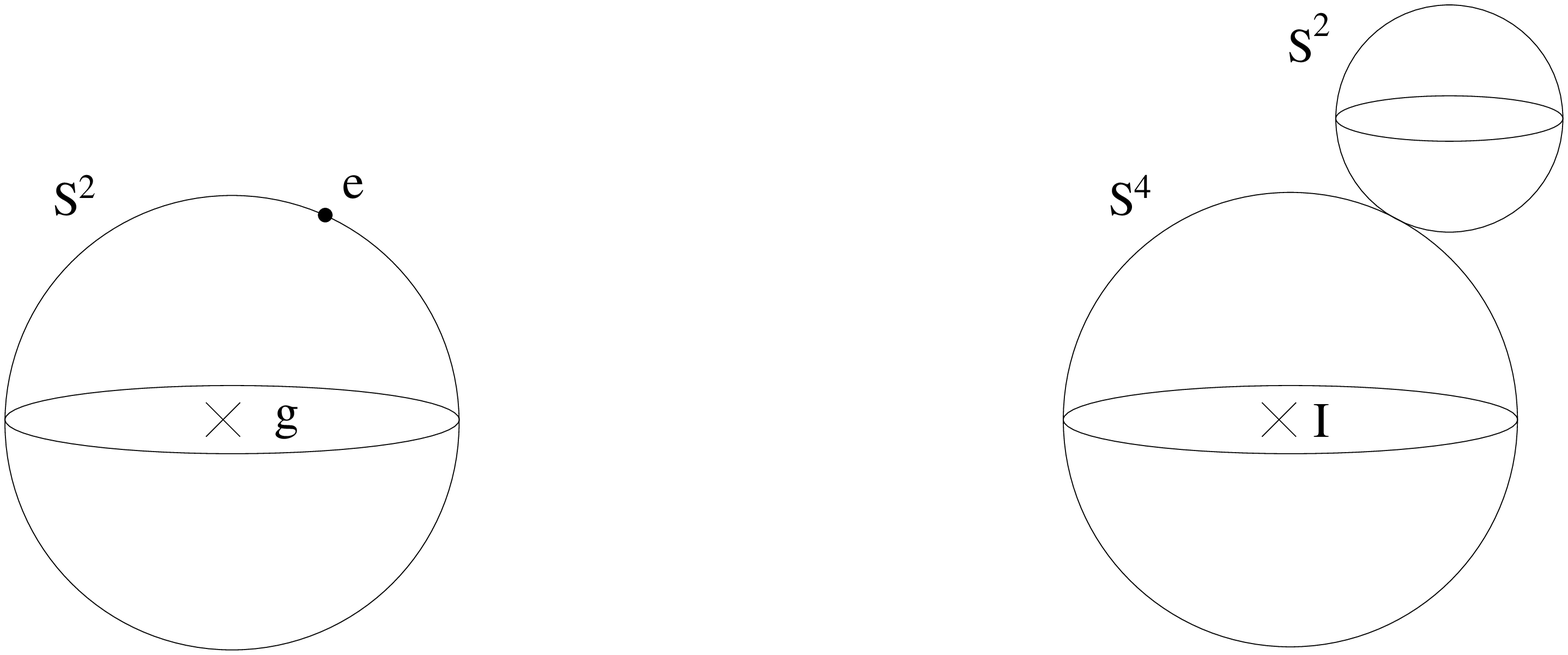} }
\caption{The 2D QHE consists of electrons $e$ on the surface of a 2D
sphere $S^2$, with a $U(1)$ magnetic monopole $g$ at its center. Similarly,
the 4D QHE can be defined on the surface of a 4D sphere
$S^4$, with a $SU(2)$ monopole $I$ at its center. In the
large $I$ limit, the $SU(2)$ isospin degree of freedom is $S^2$.
}
\label{spheres}
\end{figure}

Our last motivation to generalize QHE
comes from its geometric structure. As pointed out by Haldane\cite{haldane1983A},
a nice way to study QHE is by mapping it to the surface of a
2D sphere $S^2$, with a Dirac mangnetic monopole at its center.
(see Fig. \ref{spheres}). The Dirac quantization condition implies that the product of
the electric charge, $e$, and the magnetic charge, $g$, is quantized, {\it i.e.}
$eg=S$, where $2S$ is a integer. The number $2S+1$ is the degeneracy of the
lowest Landau level. The reason for the existence of a magnetic monopole
over $S^2$ is a coincidence between algebra and geometry. In order for the
monopole potential to be topologically non-trivial, the gauge potentials
extended from the north pole and the south pole have to match non-trivially
at the equator. Since the equator, $S^1$, and the gauge group, $U(1)$, are
isomorphic to each other, a non-trivial winding number exists.
Therefore, one may ask whether there are
other higher-dimensional spheres for which a similar monopole structure can
be defined. This naturally leads to the requirement that the equator of a
higher-dimensional sphere to be isomorphic to a mathematical group. This
coincidence occurs only for the four sphere, $S^4$, whose equator, $S^3$, is
isomorphic to the group $SU(2)$. This coincidence between algebra and
geometry leads to the first two Hopf maps, $S^3\rightarrow S^2$ and
$S^7\rightarrow S^4$.

Therefore, all three considerations -- the physical motivation of the transverse
current, the mathematical motivation of generalizing complex numbers
to quaternionic numbers and the geometric consideration of non-trivial
monopole structures -- lead to the same conclusion: A non-trivial
QHE liquid can be defined in four spatial dimensions (4D) with a
$SU(2)$ non-abelian gauge group.
Recently, Hu and I (ZH) indeed succeeded in constructing such a model for the
4D QHE\cite{Zhang2001A}. The microscopic Hamiltonian describes
a collection of $N$ fermionic particles moving on $S^4$, interacting with a
$SU(2)$ background isospin gauge potential $A_a$. It is explicitly defined
by
\begin{equation}
H=\frac{\hbar^2}{2MR^2}\sum_{a<b}\Lambda_{ab}^2
\label{4DQHE_H}
\end{equation}
where $M$ is the mass of the fermionic particle, $R$ is the radius of $S^4$, and
$\Lambda_{ab}= -i(x_a D_b - x_b D_a)$ is the gauge covariant angular momentum operator.
Here $x_a$ is the coordinate of the fermionic particle and
$D_a=\partial_a+A_a$ is the gauge invariant momentum operator. The gauge potential
$A_a$ ($a=1,2,3,4,5$) is given by
\begin{equation}
A_\mu = \frac{-i}{1+x_5}\eta_{\mu\nu}^i x_\nu I_i \ \ , \ \
A_5=0
\end{equation}
where $I_i$ are the generators of the $SU(2)$ gauge group.
An important parameter in this problem is $I$, the isospin quantum number carried by
the fermionic particle. The eigenstates and the eigenvalues of this
Hamiltonian can be solved completely, and the spectrum shares many
properties with the Landau levels in the 2D QHE problem. In particular,
when $I$ becomes large, the ground state of this problem is massively
degenerate, with the degenracy scaling like $D\sim I^3$. In order to
keep the energy levels finite in the thermodynamical limit, one is
required to take the limit $I\rightarrow\infty$ as $R\rightarrow\infty$,
such that
\begin{equation}
R^2/2I=l^2
\label{magnetic_length}
\end{equation}
is finite. $l$, called the magnetic length, defines the fundamental length
scale in this problem. It gives a natural ultraviolet cut-off in this theory,
without breaking any rotational symmetries of the underlying Hamiltonian.

While the 4D QH liquid can be elegantly defined on $S^4$, with the full
isometry group as the symmetry of the Hamiltonian, it can also be defined
on $R^4$, with more restricted symmetries. This construction has recently
been given by Elvang and Polchinski\cite{elvang2002A}.

\subsection{Properties of the model}
The 2D QH liquid has many interesting properties including incompressibility
of the quantum liquid,
fractional charge and statistics of elementary excitations, a topological
field theory description of the low energy physics, a realization of non-commutative
geometry and relativistic chiral excitations at the edge of the QH
droplet. Most of these properties also carry over to the QH liquid constructed
by ZH.
When one completely fills the massively degenerate lowest energy ground states
with fermionic particles, with filling factor $\nu\equiv N/D=1$, one obtains
an incompressible quantum
liquid, with a finite excitation gap towards all excited states.
FQH states can also be constructed for filling fractions
$\nu=1/k^3$, where $k$ is a odd integer. Explicit microscopic wave functions,
similar to Laughlin's wave function for the 2D QHE, can be constructed
for these incompressible states. The elementary excitations of the
FQH states also carry fractional charge $1/k^3$, providing
the first direct generalization of fractional charge in a higher-dimensional
quantum many-body system.

As discussed in section (\ref{topology}),
the low energy physics of the 2D QHE can be described
by a topological quantum field theory, the CSLG theory. A natural question is
whether the QH liquid constructed by ZH can be described by a
topological quantum field theory
as well. This construction has indeed been accomplished recently, by Bernevig,
Chern, Hu, Toumbas and myself\cite{Bernevig2002A}. As explained
earlier, while the underlying orbital space for our QH liquid is four
dimensional, the fermionic particles also carry a large internal isospin
degree of freedom $I$. Since $I$ scales like $R^2$, the internal space is
2D, which makes the total configuration space a six-dimensional (6D)
manifold. Therefore, our QH liquid can either be viewed as a 4D QH liquid
with a large internal $SU(2)$ isospin degrees of freedom, or equivalently, as
a 6D QH liquid without any internal degree of freedom.
The 6D manifold is $CP_3$, the complex projective space
with three complex (and therefore six real) dimensions. This manifold is
locally isomorphic to $S^4\times S^2$. The deep connection between the four
sphere $S^4$ and the complex manifold $CP_3$ was first introduced to physics through
the twistor program of Penrose\cite{penrose1972A} and has been exploited extensively
in the mathematical literature. Sparling\cite{sparling2002A} has recently
pointed out the close connection between the twistor theory and the 4D QHE.
Our recent work shows that the low energy effective field
theory of our QH liquid is given by an abelian CS theory in $6+1$ dimensions
\begin{eqnarray}
S=\nu \int dt d^6x A \wedge dA \wedge dA \wedge dA
\label{CS_CP3}
\end{eqnarray}
where $A$ is an abelian $U(1)$ gauge field over the total configuration space
$CP_3$, and $\nu$ is the filling factor.
This theory can also be dimensionally reduced to a $SU(2)$ non-abelian CS theory
in $4+1$ dimensions, given by
\begin{equation}
S=\frac{4\pi\nu}{3} \int dt d^4x
Tr\left({\bf A} \wedge d{{\bf A}} \wedge d{\bf A} - {3i \over
2}{\bf A} \wedge {\bf A} \wedge {\bf A} \wedge d{{\bf A}}- {3
\over 5}{\bf A} \wedge {\bf A} \wedge {\bf A} \wedge {\bf A}\wedge
{\bf A} \right)
\label{CS_S4}
\end{equation}
where ${\bf A}$ is a $SU(2)$ matrix-valued gauge field over the orbital space
$S^4$. The precise equivalence of these two models parallels the two equivalent
views of our QH liquid mentioned earlier.

An interesting property which arises from this field theory is
the concept of duality. As discussed in section (\ref{topology}),
there is a natural particle-flux
duality in the 2D QHE problem: An electron can be represented as a boson with
an odd number of flux quanta attached to it. In the new QH liquid, there are other
extended objects, namely 2-branes and 4-branes besides the basic fermionic
particle, which can be viewed as a 0-brane. Each one of these extended objects
is dual to a generalized flux, according to the following table:
\begin{center}
\begin{tabular}{ccc}
Particle & $\Longleftrightarrow$ &  6-flux  \\
Membrane & $\Longleftrightarrow$ &  4-flux  \\
4-brane  & $\Longleftrightarrow$ &  2-flux  \\
\end{tabular}
\end{center}
In the 2D QH problem, the Laughlin quasi-particles obey fractional statistics in 2+1
dimensions. It is natural to ask how fractional statistics generalize in
our QH liquid. It turns out that the concept of fractional statistics of point
particles can not be generalized to higher dimensions, but fractional statistics
for extended objects exist in higher dimensions\cite{wu1988A,tze1989A}.
In our case, 2-branes have
non-trivial statistical interactions which generalizes statistical interactions
of Laughlin quasi-particles.

Extended objects like D-branes have been studied extensively in string theory,
however, a full quantum theory describing their interactions
still needs to be developed. The advantage of our approach is that the underlying
microscopic quantum physics is completely specified. Since the extended topological
objects emerge naturally from the underlying microscopic physics, there is hope that
a full quantum theory can be developed in this case.

The study of 4DQHE is partially motivated by the possibility of emergent relativistic
behavior in $3+1$ dimensions. There are several ways to see the connection.
First of all, the eigenstates and the eigenfunctions of the Hamiltonian (\ref{4DQHE_H})
have a natural interpretation in terms of the 4D Euclidean quantum field theory.
If we consider a Euclidean quantum field theory as obtained from a Wick
rotation of a $3+1$ dimensional compactified Minkowskian quantum field theory, one is naturally
lead to consider the eigenvalues and the eigenfunctions of the Euclidean
Dirac, Maxwell and Einstein operators on $S^4$. It turns out that the these
eigenvalues and eigenfunctions coincide exactly with the eigenvalues and
eigenfunctions of the 4DQHE Hamiltonian (\ref{4DQHE_H}), where the spins of the relativistic
particles are identified with the isospin quantum number, $I$.
The eigenvalue problems of the Dirac, Maxwell and Einstein operators can be
directly identified with the Hamiltonian eigenvalue problems for
$I=1/2,1$ and $2$. We mentioned earlier that the underlying fermionic
particles constituting our QH liquid have high isospin quantum numbers.
However, collective excitations of this QH liquid, which are formed
as composite particles, can have low isospin quantum numbers. It is therefore
tempting to identify the collective excitations of the QH liquid with the relativistic
particles we are familiar with. However, this equivalence is only established
in Euclidean space. In order to consider the relationship to Minkowski space,
we are naturally lead to the excitations at the boundary, or the edge of
our QH liquid.

Let us first review the collective excitations at the edge of a 2D QH liquid.
The 2D QH liquid can be confined by a one-body confining potential $V$. A density
excitation is created by removing a particle from the QH liquid and placing
it outside of the QH liquid. This way, we have created a particle-hole excitation.
If the particle-hole pair moves along a direction parallel to the edge, with a
center of mass momentum $q_x$, the Lorentz force due to the magnetic field acts
oppositely on the particle-hole pair, and tries to stretch the pair in the
direction perpendicular to the edge. This Lorentz force is balanced by the
electrostatic attraction due to the force of the confining potential. Therefore,
a unique dipole moment, or a finite separation $y$ of the particle hole pair,
is obtained in terms of $q_x$:
\begin{equation}
y=l^2 q_x
\label{dipole}
\end{equation}
On the other hand, the energy of the dipole pair is simply given by
$E=V' y$. Here $V'$ is the derivative of the potential evaluated at the edge.
Therefore, we obtain a relativistic dispersion relation for the
dipole pair
\begin{equation}
E=V' y=l^2 V' q_x
\label{dipole_dispersion}
\end{equation}
with the speed of light given by $c=l^2 V'$. Since the cross product of the
gradient of the potential and the magnetic field selects a unique direction
along the edge, the excitation is also chiral. In this problem, it can also
be shown that not only the dispersion, but also the full interaction is
relativistic in the low energy limit. Therefore, the physics at the
edge of a 2D QH liquid provides another example of emergent relativistic
behavior\cite{wen1990A,stone1990A}.

The physics of the edge excitations of a 2D QH liquid {\it partially} carries over
to our 4D QH liquid\cite{Zhang2001A,hu2002A,elvang2002A}.
Here we can also introduce a confining potential, say
around the north pole of $S^4$, and construct a droplet of the QH fluid.
Since our QH liquid is incompressible, the only low energy excitations
are the volume preserving shape distortions at the surface. These surface
waves can be formed from the particle-hole excitations similar to the ones
we described for the 2D QH liquid. A natural speed of light can be introduced,
and is given by $c=l^2 V'$. Since our underlying particles carry a large
isospin, $I$, the bosonic composite particle-hole excitations carry all
isospins, ranging from $0$ to $2I$. The underlying
fermionic particles have a strong coupling between their orbital and isospin
degrees of freedom. This coupling translates into a relativistic spin-orbital
coupling of the bosonic collective excitations. Therefore, excitations
with $I=0,1,2$ obey the {\it free} relativistic Klein-Gordon, Maxwell and Einstein
equations. This is an encouraging sign that one might be able to construct
an emergent relativistic quantum field theory from the boundary excitations
of our 4D QH liquid.

However, there are also many complications which are not yet fully understood
in our approach. The most fundamental problem is that particles of our 4D
QH liquid carry a large internal isospin, which makes the problem effectively
a 6D one. This is the basic reason for the proliferation of
higher-spin particles in our theory, an ``embarrassment
of riches". In addition, there is an incoherent fermionic continuum
besides the bosonic collective modes. All these problems can only be
addressed when one studies the effects of the interaction carefully.
In fact, single particle states in the lowest-Landau-level (LLL)
have the full symmetry of $SU(4)$, which is the isometry group of the six dimensional
$CP_3$ manifold. In order to make the problem truly 4D,
one needs to introduce interactions which breaks the $SU(4)$ symmetry to
a $SO(5)$ symmetry, the isometry group of $S^4$. This is indeed
possible. $SO(5)$ is isomorphic to the group $Sp(4)$. $Sp(4)$ differs
from $SU(4)$ by an additional reality condition, implemented through a
charge conjugation matrix $R$. Therefore, any interactions which involve
this $R$ matrix would break the symmetry from $SU(4)$ to $SO(5)$, and
the geometry of $S^4$ would emerge naturally. In the strong coupling
limit, low energy excitations are not particles but membranes. This
reduces the entropy at the edge from $R^3\times R^2$ to $R^3$, and
is the first step towards solving the problem of ``embarrassment
of riches".

\subsection{Space, time and the quantum}
The 2D QH problem gives a precise mathematical realization of the concept of
non-commutative geometry\cite{douglas2001A}.
 In the limit of high magnetic field, we can take
the limit of $m\rightarrow 0$, so that all higher Landau levels are projected
out of the spectrum. In this limit, the equation of motion for a charged particle
in an uniform magnetic field $B$ and a scalar potential $V(x,y)$ is given by
\begin{equation}
\dot{x}  = l^2 \frac{\partial V}{\partial y}\ \ , \ \
\dot{y}  = -l^2 \frac{\partial V}{\partial x}
\label{EOM}
\end{equation}
We notice that the equations for $x$ and $y$ look exactly like the Hamilton
equations of motion for $p$ and $q$. Therefore, this equation of motion can
be derived as quantum Heisenberg equations of motion if we postulate a
similar commutation relation:
\begin{equation}
[x,y]=il^2
\label{XY}
\end{equation}
Therefore, the 2D QHE provides a physical realization of the mathematical
concept of non-commutative geometry, in which different spatial components
do not commute. Early in the development of quantum field theory, this
feature has been suggested as a way to cut off the ultraviolet divergences
of quantum field theory. In quantum mechanics, the non-commutativity of
$q$ and $p$ leads to the Heisenberg uncertainty principle and resolves the
classical catastrophe of an electron falling towards the atomic nucleus.
Similarly, non-commutativity of space and time could cut off the ultraviolet
space-time fluctuations in quantum gravity\cite{douglas2001A}.
However, the problem is that
equation (\ref{XY}) can not be easily generalized to higher dimensions, since one needs
to pick some fixed pairs of non-commuting coordinates. Our QH liquid provides
a physical realization of non-commutative geometry in four dimensions.
The generalization of equation (\ref{XY}) becomes
\begin{eqnarray}
[X_\mu,X_\nu]= 4 i l^2 \eta_{\mu\nu}^i n_i
\label{non-commutation}
\end{eqnarray}
where $X_\mu$'s are the four spatial coordinates and $n_i$ is the isospin coordinate
of a particle. This structure of non-commutative geometry is invariant
under a combined rotation of space and isospin and treats all these
coordinates on equal footing. It is tempting to identify $l$ in equation
(\ref{non-commutation}) as the Planck length, which provides the fundamental cutoff of the length
scale according to the quantization rule (\ref{non-commutation}). In our theory, however, we
know what lies beyond the Planck length: the degrees of freedom are those
associated with the higher Landau levels of the Hamiltonian (\ref{4DQHE_H}).

At this point, it would be useful to discuss the
possible implications of (\ref{non-commutation})
on the quantum structure of space-time. In the 4D QH liquid, there is
no concept of time. Since all eigenstates in the LLL are degenerate, there is
no energy difference which can be used to measure time according to the
quantum relation $\Delta t =\hbar/\Delta E$. However, at the boundary of the
4D QH liquid, an energy difference is introduced through the confining
potential. The left hand side of equation (\ref{non-commutation})
involves four coordinates.
Three of them are the spatial coordinates parallel to the boundary. The
fourth coordinate, perpendicular to the boundary, measures the energy
difference, and therefore measures time. The commutator among these
coordinates implies a quantization procedure. The right hand side of this
equation involves the Planck length and the spin. Therefore, this simple
equation seems to unify all the fundamental physical concepts: space,
time, the quantum, the Planck length and spin in a simple and
elegant fashion. It would be nice to use it as a basis to construct a
fundamental physical theory.

\subsection{Magic liquids, magic dimensions, magic convergence?}
So far our philosophical point of view and our model seem to be
drastically different from the approach typical of string theory.
However, after the discovery of the new QH liquid, a surprising pattern
starts to emerge. Soon after the construction of the new 4D QH liquid,
Fabinger\cite{fabinger2001A} found that it
could be implemented as certain solutions in string theory. Moreover,
close examination of this pattern reveals remarkable
mathematical similarities not only between these two approaches, but also
with other fundamental ideas in algebra, geometry, supersymmetry and
the twistor program on quantum space time. The following table
summarizes the connections.

\vspace{.2in}
\begin{center}
\begin{tabular}{|c|c|c|c|c|}    \hline
Division Algebras: & Real Namubers & Complex Numbers   &
Quaternions & Octonions\\  \hline Hopf maps: & $S^1 \rightarrow
S^1$ & $S^3 \rightarrow S^2$   & $S^7 \rightarrow S^4$ &
$S^{15}\rightarrow S^8$\\  \hline QH liquids: &Luttinger liquid? &
Laughlin liquid  & ZH liquid & ?\\  \hline
Random matrix ensembles: &Orthogonal& Unitary & Symplectic & ?\\  \hline
Fractional statistics: &? & particles  & membranes & ?\\  \hline
Geometric phase: &$Z_2$ & $U(1)$  & $SU(2)$ & ?\\  \hline
Non-commutative geometry: &? & $[X_i,X_j]= i l^2 \epsilon_{ij}$  &
$[X_\mu,X_\nu]= 4 i l^2 \eta_{\mu\nu}^i n_i$ & ?\\  \hline Twistor
transformation: & $SO(2,1)=SL(2,{\bf R})$ & $SO(3,1)=SL(2,{\bf
C})$ & $SO(5,1)=SL(2,{\bf H})$  & $SO(9,1)=SL(2,{\bf O})$\\
\hline $N=1$ SUSY Yang-Mills: & $d=2+1$ & $d=3+1$  & $d=5+1$ &
$d=9+1$\\  \hline Green-Schwarz Superstring & $d=2+1$ & $d=3+1$  &
$d=5+1$ & $d=9+1$\\  \hline
\end{tabular}
\end{center}
\vspace{.2in}

The construction of the twistor transformation, the $N=1$ supersymmetric
Yang-Mills theory and the Green-Schwarz superstring rely on certain
identities of the Dirac Gamma matrices, which work only in certain
magic dimensions. In these dimensions, there is an exact equivalence
between the Lorentz group and the special linear tranformations of
the real, complex, quaternionic and octonic numbers. Our work shows
that QH liquids work only in certain magic
dimensions exactly related to the division algebras as well!
In fact the {\it transverse} dimensions $((D+1)-2)$ of these relativistic field theories
match exactly with the {\it spatial} dimensions of the quantum liquids.
The missing entries in this table strongly suggests that an octonionic version
of the QH liquid should exist and may be deeply related to the
superstring theory in $d=9+1$.
QH liquids exist only in magic dimensions, have membranes and look like a matrix
theory. They may be mysteriously related to the M theory after all!

\section{Conclusion}
\label{conclusion}

Fundamental physics is faced with historically unprecedented
challenges. Ever since the time of Galileo, experiments have been
the stepping stones in our intellectual quest for the fundamental
laws of Nature. With our feet firmly on the ground, there is no
summit too high to reach. However, the situation is drastically
different in the present day. We are faced with a gap of 16 orders
of magnitude between the energy of our experimental capabilities
and the summit of Mount Planck. Without experiments, we face the
impossible mission of climbing up a waterfall!

\begin{figure}[h]
\centerline{\epsfysize=8cm \epsfbox{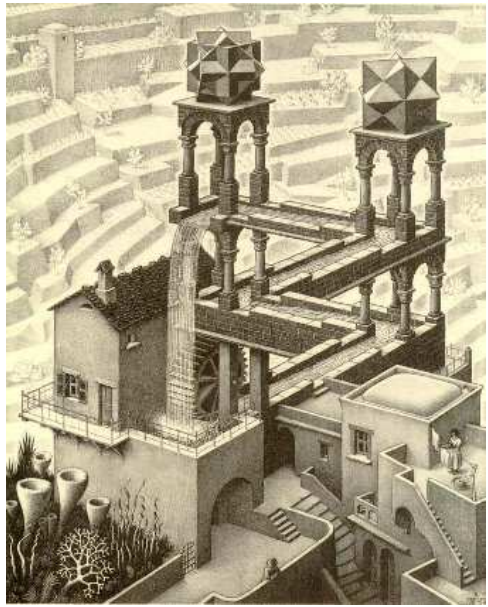} }
\caption{Esher's waterfall: an alternate passage to Mount Planck?
}
\label{waterfall}
\end{figure}

But maybe there is an alternate passage to Mount Planck. The logical
structure of physics may not be a simple one-dimensional line, but
rather has a multiply connected or braided topology, very much like
Esher's famous {\it Waterfall}.
Instead of going up in energy, we can move down in energy!
Atoms, molecules and quantum liquids are made of elementary
particles at very high energies. But at low energies, they interact
strongly with each other to form quasi-particles, which look very
much like the elementary particles themselves! Over the past forty years,
we have learned that the strong correlation of these matter degrees
of freedom does not lead to ugliness and chaos, but rather
to extraordinary beauty and simplicity. The precision of flux quantization,
Josephson frequency and quantized Hall conductance are not
properties of the basic constituents of matter, but rather are emergent
properties of their collective behavior. Therefore, by exploring the
connection between elementary particle and condensed matter physics,
we can use experiments performed at low energies to understand the
physics at high energies. By carrying out the profound implications of
these experiments to their necessary logical conclusions, we may
learn about the ultimate mysteries of our universe.

Throughout John Wheeler's life, he tackled the big questions of the universe
with an unorthodox vision and a poetic flair.
Lacking John's eloquence, I simply conclude this tribute to him by
reciting William Blake's timeless lines:

\begin{center}
\it{
To see a World in a Grain of Sand

And a Heaven in a Wild Flower,

Hold Infinity in the palm of your hand

And Eternity in an hour.}
\end{center}

\section*{ACKNOWLEDGMENTS}

I would like to thank A. Bernevig, C.H. Chern, J.P. Hu, R. B.
Laughlin, J. Polchinski, P. SanGiorgio, L. Smolin, L. Susskind, N.
Toumbas and G. Volovik for stimulating discussions. This work is
supported by the National Science Foundation under grant number
DMR-9814289.

\bibliographystyle{apsrmp}
\bibliography{wheeler}

\end{document}